# FRONT-END ELECTRONICS CONFIGURATION SYSTEM FOR CMS

P. Gras, CERN, Switzerland, University of Karlsruhe/IEKP, Germany
W. Funk, CERN, Geneva, Switzerland
L. Gross, D. Vintache, IReS, Strasbourg, France
F. Drouhin, UHA, Mulhouse, France

*Abstract*

The four LHC experiments at CERN have decided to use a commercial SCADA (Supervisory Control And Data Acquisition) product for the supervision of their DCS (Detector Control System). The selected SCADA, which is therefore used for the CMS DCS, is PVSS II from the company ETM. This SCADA has its own database, which is suitable for storing conventional controls data such as voltages, temperatures and pressures. In addition, calibration data and FE (Front-End) electronics configuration need to be stored. The amount of these data is too large to be stored in the SCADA database [1]. Therefore an external database will be used for managing such data. However, this database should be completely integrated into the SCADA framework, it should be accessible from the SCADA and the SCADA features, e.g. alarming, logging should be benefited from. For prototyping, Oracle 8i was selected as the external database manager. The development of the control system for calibration constants and FE electronics configuration has been done in close collaboration with the CMS tracker group and JCOP (Joint COntrols Project)[1].

## 1 INTRODUCTION

The supervision of the CMS detector control system (DCS) has to control classical "slow control" items such as power supplies, gas systems, etc. as well as FE (Front End) configuration, which comprises a lot of different parameters.

The supervision level has to provide hardware access, alarming, archiving, scripting, graphical user interface and logging features. In the following the prototype system developed for the tracker of CMS will be described.

---

[1] The four LHC experiments and the CERN IT/CO group has merged their efforts to build the experiments controls systems and set up the JCOP at the end of December, 1997 for this purpose.

## 2 DESCRIPTION OF THE FRONT-END CONFIGURATION CONTROL SYSTEM

PVSS II, a commercial SCADA product from the Austrian company ETM was used for the supervision level. Because the amount of data for the FE configuration is quite big, an external database

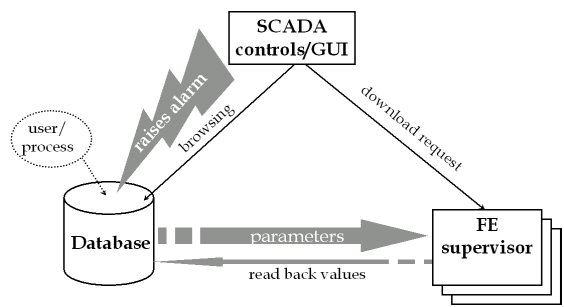

Figure 1: FE configuration mechanism.

(Oracle 8i) was used for the storage of the FE configuration data itself.

### 2.1. Configuration mechanism

The FE configuration system is comprised of 3 actors as illustrated in Figure 1: the database, the SCADA and FE supervisor(s). The database stores the parameters, the SCADA controls the operation and

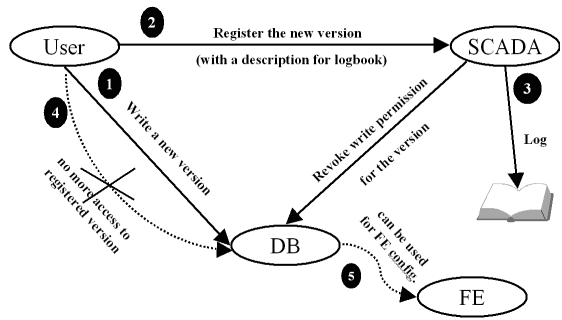

Figure 2: Access control registration mechanism.

provides the user interface, the FE supervisor(s)

accesses the FE. The SCADA system as well as the database can be distributed over many PCs on different platforms. In order to transfer the data in parallel to the electronics, the system can have several FE supervisors.

On user request or during an automatic procedure (e.g. start of run, error recovery…) the SCADA sends a download command to the FE supervisors. Then the FE supervisors fetch the data from the database and download them into the FE electronics. It is possible to ask each FE supervisor to read back the configuration from the electronics and send it back to the database. Then, if some values are outside limits, an alarm summarizing the differences will be sent to the SCADA.

## 2.2. SCADA library and user interface

For the user interface, PVSS II panels have been developed: for version creation, for version registration, for database browsing, etc. These panels can be used as complex widgets.

The database-browsing panel called DBNav is a generic user interface for Oracle 8i databases. It is able to discover itself the structure of the database and display it in a tree.

These panels are based on two underlying PVSS II script libraries, which can be used directly to develop custom scripts or panels.

## 2.3. Access control

In order to keep the history of the data used for a configuration, versioning of the stored parameters and a registration mechanism have been developed. FE parameters, and also calibration constants, may be calculated by some process, which is independent of the SCADA. Before a configuration set may be used for a run, it should be registered using the SCADA. At registration time, the SCADA logs the description of the new configuration, and revokes the write permission on the registered configuration set. This write permission revocation is done using the database access control. In this way, all versions used once for configuration will be kept unchanged for later analysis. Figure 2 describes this registration mechanism.

## 2.4. Database model

Each electronic device type is represented by a table. This table contains the device parameters. In the example of the CMS Tracker described in section 3.2, the database table named APV contains all the parameters of the APV readout chips. A device can be part of a higher-level device (e.g. a chip is part of a board). Such membership relation is specified in the database by a standard relational database "references constraint"[2] between the device and the subsystem. A "controlled by" relationship or any N-to-1 relationship is represented by such a constraint.

The parameter versioning is taken care of by a specific table, typically called "version", which contains the list of all available versions. A version is identified by two numbers: the major and the minor version ids. The version table is composed of at least five columns: one for the major id, one for the minor id, one for the version creation date, one for the description and one which specifies if the version has been registered. Each row of device type tables contains the values of one device for a specific parameter version. The row includes the version ids, which refer ("references constraint") to the version table. Actually if a device contains versioned parameters and version-independent parameters, the device type table can be split into two tables: one for the versioned parameters and one for the version-independent parameters.

Finally, in order to use the alarm mechanism, the device type table must have a "device_type" column which specifies if the row contains the set value, the minimum allowed value or the maximum allowed value.

The "value_type" column and the version table are optional and are not needed if alarming or versioning is not required.

## 3 FE CONFIGURATION SYSTEM USAGE

## 3.1. Electronics-specific part

Two parts are specific to the front-end. The first part is the database content: typically each device will have a table, which will contain its parameters. A general database scheme is given as a template.

The other specific part is the "driver". The "driver" is the part that fetches the data from the database and downloads it to the front-end electronics. It receives commands from the SCADA[3]. This part needs to know how to access the specific front-electronics hardware. It can get the data from the database using a standard interface like JDBC, as it has been done for the Tracker (see next part) or with more Oracle-specific interface like OCI or Pro*C or in XML format (provided as standard by Oracle 8i).

---

[2] In relational database jargon the constraints are the rules that are defined in order to keep the database consistent. Each time the database is altered, the database manager checks that these rules are not violated.
[3] A simple interface to receive this command is provided in C/C++ and in Java.

## 3.2. Use of the FE configuration system for the CMS tracker readout electronics

The Tracker FE has in its final design about 80,000 APV readout chips of which each has about 20 parameters. Therefore each version of parameters will contain several Mbytes. With the expected number of versions we arrive at the order of GBytes.

The Tracker is organized in modules. Each module has 2 to 6 APVs, 1 PLL[4] chip and 1 channel multiplexer, called an APVMUX.

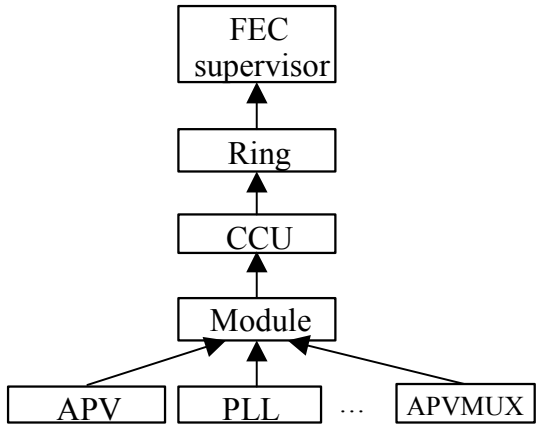

Figure 3: FEC electronics control hierarchy.

Chips, called CCUs, control the APVs. CCUs[5] communicate through a Token Ring controlled by a FEC[6][4] board. In current prototypes the FECs are PCI cards hosted by a PC. The hierarchy of all FE electronics is shown in figure 3.

This hierarchy is reflected in the database through the "references constraints" as described in Section 2.4. For each item in Figure 3 one table is defined in the database.

The setup has been used successfully in the tracker beam test, which took place at CERN in October 2001. Figure 4 shows the FE configuration in the beam test DCS context. During this beam test, PVSS II was also controlling a HV power supply and was monitoring humidity and temperatures on the detector. A PLC was interlocking the high voltage depending on the detector temperature. On interlock the PLC was notifying PVSS II in order to generate an alarm. An electronics logbook using Oracle database with a user interface in PVSS II and a web interface has also been developed for this beam test. Finally a communication between the DAQ and SCADA has been implemented in order to synchronize them.

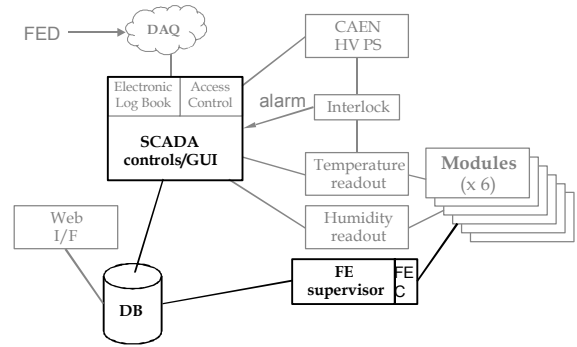

Figure 4: Tracker DCS setup. The FE electronics configuration system is represented in bold. The SCADA communicates with the DAQ, which receives data from the FEDs (Front End Drivers). For simplification the CCU ring between the FEC and the modules is not shown.

## 4 CONCLUSIONS

The FE electronics configuration system, which has been developed and applied to the CMS Tracker, has been designed in a sufficiently generic manner to be used in other detectors. A prototype using this system for the CMS ECAL detector is under development. During a beam test it has been proven that this system fulfills the principal requirements. In the future some more effort will be needed to measure the performance and to tune it. Tests will also be undertaken to verify the scalability of the system.

## ACKNOWLEDGEMENTS

We are deeply indebted towards the CERN Oracle support group for their competent assistance in setting up and running our database.

---

[4] Phase Locked Loop
[5] Communication and Control Unit.
[6] Front-End Controller